\documentclass[twocolumn,english,prb,reprint,superscriptaddress]{revtex4-2}
\usepackage{amsmath}
\usepackage[utf8]{inputenc}
\usepackage{txfonts}
\usepackage{amssymb}
\usepackage{graphicx}
\usepackage{xcolor}
\usepackage[colorlinks=true, urlcolor=blue, linkcolor=blue, citecolor=blue]{hyperref}

\makeatletter
\usepackage{babel}

\def\eV{{\,\mathrm{eV}}}
\def\meV{{\,\mathrm{meV}}}
\def\K{{\,\mathrm{K}}}
\newcommand{\cms}{\mathrm{{cm}^{2}~s^{-1}}}
\newcommand{\Scm}{\mathrm{S\cdot{cm}^{-1}}}
\newcommand{\SCO}{\mathrm{{Sr}_{2}{Co}_{2}{O}_{5}}}

\makeatother

\begin{document}
\title{Fast Lithium Ion Diffusion in Brownmillerite $\mathrm{Li}_{x}\SCO$}
\author{Xin Chen}
\affiliation{School of Physical Science and Technology, Soochow University, Suzhou
215006, China}
\author{Xixiang Zhang}
\affiliation{Physical Science and Engineering Division, King Abdullah University of Science and Technology (KAUST), Thuwal 23955-6900, Kingdom of Saudi Arabia}
\author{Jie-Xiang Yu}
\email{jxyu@suda.edu.cn}
\affiliation{School of Physical Science and Technology, Soochow University, Suzhou
215006, China}
\affiliation{Department of Physics and Astronomy, University of New Hampshire,
Durham, New Hampshire 03824, USA}
\author{Jiadong Zang}
\email{jiadong.zang@unh.edu}
\affiliation{Department of Physics and Astronomy, University of New Hampshire,
Durham, New Hampshire 03824, USA}

\begin{abstract}
Transition metal oxides not only exhibits novel magnetic properties, but also provides outstanding ionic transports. Ionic conductors have great potential for interesting tunable physical properties via ionic liquid gating and novel energy storage applications such as all-solid-state lithium batteries.
In particular, low migration barriers and high hopping attempt frequency are the keys to achieve fast ion diffusion in solids. 
Taking advantage of the oxygen-vacancy channel in $\mathrm{Li}_{x}\SCO$, we show that migration barriers of lithium ion are as small as $0.28\sim0.17~\eV$ depending on the lithium concentration rates. 
Our first-principles calculation also investigated hopping attempt frequency and concluded the room temperature ionic diffusivity and ion conductivity is high as ${10}^{-7}\sim{10}^{-6}~\cms$ and ${10}^{-3}\sim{10}^{-2}~\Scm$ respectively, which outperform most of perovskite-type, garnet-type and sulfide Li-ion solid-state electrolytes. 
This work proves $\mathrm{Li}_{x}\SCO$ as a promising super-ionic conductor.  
\end{abstract}

\maketitle

\section{Introduction}

Ionic diffusion in solids has played a key role in not only manipulating
many interesting physical properties in ion-electron-lattice-coupled
systems via ionic liquid gating (ILG)\cite{Tan_2018,Lu_2017,Li_2020}, but
also great potential applications in energy storage such as all-solid-state
lithium batteries in which solid electrolytes
with both high Li-ion conductivity and low electron conductivity\cite{Wang_2015,Aksyonov_2023, Wu_2023}. 
To achieve highly efficient ionic diffusion, 
high ion conductivity or high ionic diffusivity is required. 
According to Nernst-Einstein equation\cite{Kutner_1981,Park_2010,Aksyonov_2023,Sotoudeh_2023}, the relationship between ionic
diffusivity $D$ and ion conductivity $\sigma_{i}$ in solids can
be given by 
\begin{equation}
\sigma_{i} = q^{2}nD\beta
\label{eq:nernst-einstein}
\end{equation}
where $q$ is the electric charge of conducting ion, $n$ is the
ionic carrier concentration and $\beta=(k_{B}T)^{-1}$ is the the
Boltzmann factor, the inverse of the product of Boltzmann constant
and temperature. 
Meanwhile, under kinetically ideal conditions, the
diffusivity can also be described by ion hopping through a pathway
with a hopping frequency 
derived by Vineyard in the classical limit\cite{Vineyard_1957,Aksyonov_2023,Sotoudeh_2023}:
 
\begin{equation}
D = a^{2}\nu^{*}\exp\left(-E_{a}\beta\right)
\label{eq:d-model}
\end{equation}
where $a$ is the hopping distance between two neighbor stable sites, 
$\nu^{*}$ is the hopping attempt frequency  
and $E_{a}$ is the activation energy of ions or the migration
barriers during the diffusion. 
To this end, the high hopping attempt
frequency and the low migration barriers is crucial for fast ionic diffusion.

Perovskite-type oxide systems are promising candidates for all-solid-state lithium batteries due to their well ordered
diffusion channels and the high Young's modulus\cite{Lu_2021a}. 
Furthermore, the
desirable combination of the complex electron-lattice-spin coupling,
the strongly correlated $d-$electrons and the multivalent transition
metal ions in transition metal oxides brings about novel electronic
and magnetic properties\cite{Imada_1998,Dagotto_2005,Ngai_2014}.
Similar to the perovskite-type oxides, 
the recent study on brownmillerite $\SCO$
demonstrated that ILG could induce tri-state phase transformation
from $\SCO$ to perovskite $\mathrm{SrCoO}_{3}$
and $\mathrm{H}_{2}\SCO$ by the insertion of oxygen
anions and hydrogen cations respectively\cite{Lu_2017}.
Han \textit{et al.}\cite{Han_2022} also reported that the direction of the oxygen-vacancy channels can be well controlled.
The well-ordered\cite{Munoz_2008, Mitra_2014} and controllable oxygen-vacancy
channels in brownmillerite $\SCO$ provide favorable
conditions for fast ionic diffusion and storage of ions, bringing
about applications in fuel cells and rechargeable batteries.

In this paper, we systemically studied the diffusion of $\mathrm{Li}^{+}$
ion in brownmillerite $\SCO$ based on first-principles calculations. 
The injected $\mathrm{Li}^{+}$ cations can be stabilized inside the oxygen-vacancy channels to form $\mathrm{Li}_{x}\SCO\left(x=0.0\sim1.0\right)$.
After confirming the chemical stability, 
we obtained the migration barrier of $\mathrm{Li}^{+}$ cations 
along the oxygen-vacancy channels as $0.28\sim0.17\eV$ $\mathrm{Li}^{+}$ depending on $\mathrm{Li}^{+}$ concentration rates. 
The low migration barrier is caused by the multi-bonding property of Li-O bonds, and is lower than most of the Perovskite-type Li-ion solid electrolytes. 
The corresponding diffusivity and conductivity at room temperature are obtained as $10^{-7} \sim 10^{-6}~\cms$ and $10^{-3} \sim 10^{-2}~\Scm$ respectively. Such high ionic diffusivity and ion conductivity can be considered as a super-ionic conductor\cite{He_2017}.

\section{Methodology}

We performed density-functional theory (DFT) based calculations with
projector augmented wave (PAW) pseudopotentials\cite{PAW_1994,PAW_1999}
implemented in the Vienna ab initio simulation (VASP) package\cite{Kresse_1996_CMS,Kresse_1996_PRB}.
The generalized gradient approximation in Perdew, Burke, and Ernzerhof
formation\cite{PBE} was used as the exchange-correlation energy. 
The PAW pseudopotential set one $2s$ electron on each Li atom, 
eight $3d$ and one $4s$ electrons on each Co atom, 
two $4s$, six $4p$ and two $5s$ electrons on each Sr atom, 
and two $2s$ and four $2p$ electrons on each O atom as the valence electrons.
We employed the Hubbard $U$ method in the Liechtenstein implementation\cite{Liechtenstein_1995a} of $U=5.0\eV$, $J=0.9\eV$ 
on Co($3d$) orbitals to include the on-site strong-correlation effects of the localized $3d$ electrons, 
by consulting the theoretical studies on the hydrogen-intercalated $\SCO$\cite{Lu_2017,Lu_2022}.
An energy cutoff of 600 eV was used for the plane-wave expansion
throughout the calculations. 
The $k$-points were sampled on a $10\times10\times4$
Monkhorst-Pack mesh in the Brillouin zone of the unit cell of $\SCO$
containing eight Co, eight Sr and twenty O atoms. 
All calculation results are converged when the total energy change and the band-structure-energy change between two self-consistent field steps are both less than $10^{-8}\eV$.
For structural relaxations
we relaxed the atoms until the Hellmann-Feynman forces were less than
$1\meV/$\AA. 

In order to find the diffusion path and the corresponding
migration barriers for Li diffusion in $\mathrm{Li}_{x}\SCO$,
we performed climbing image nudged elastic band (cl-NEB) calculations\cite{NEB_2000,NEB_2000a}
to find the saddle point and the minimum energy path between two stable
local minimum. 
Eight images were employed and the force convergence
is down to $0.02\eV/$\AA. 
Phonon modes of the diffused $\mathrm{Li}^{+}$
ions were obtained by using the finite displacement method\cite{Finite_Phonon}
implemented in the Phonopy package\cite{Phonopy}.

\section{Revisit Structural, Electronic and Magnetic Properties of $\SCO$}

\begin{figure}
\includegraphics[width=1\columnwidth]{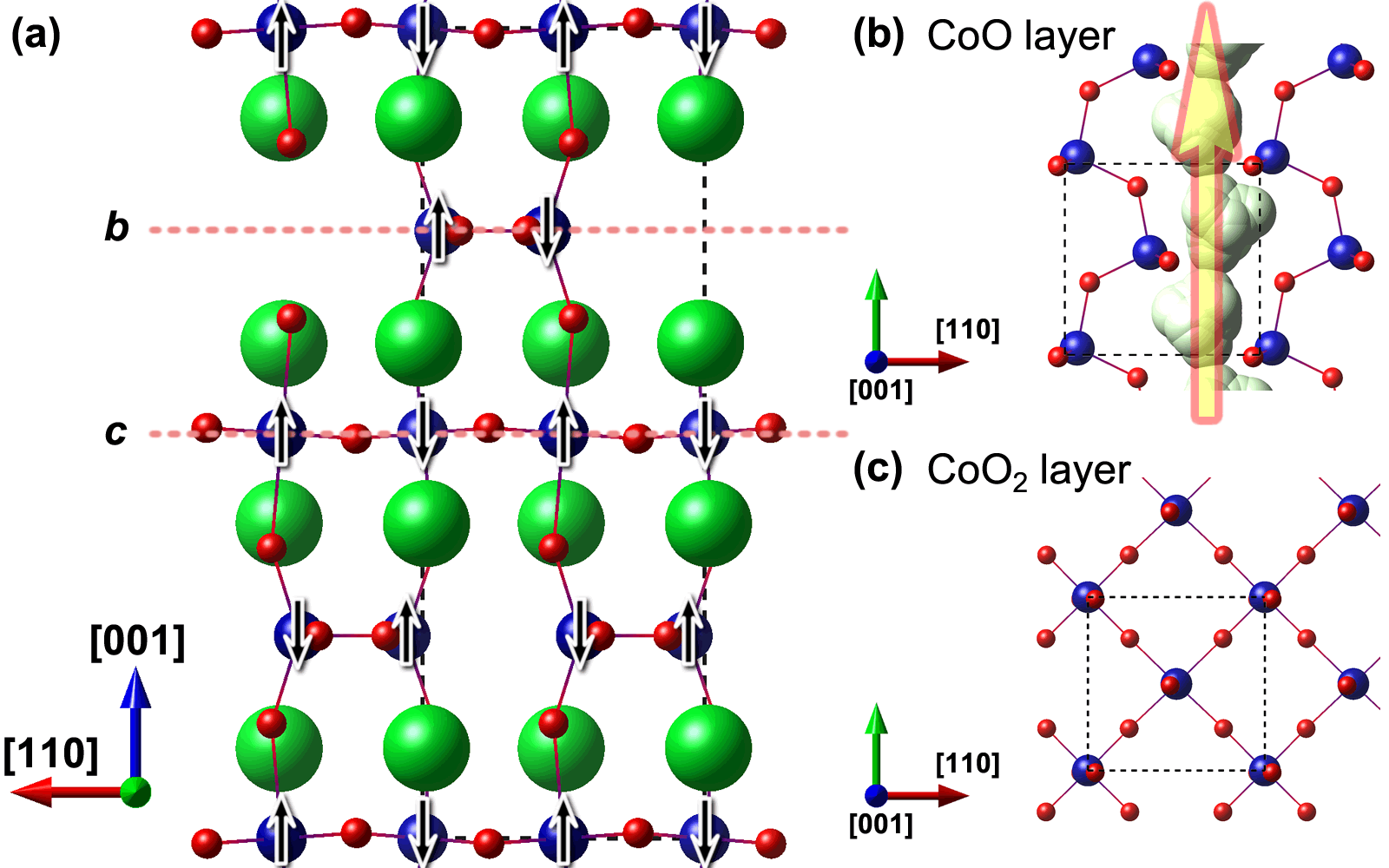}
\caption{Crystal structure of $\SCO$ 
in one unit cell.
(a) The side view along $\left[\bar{1}10\right]$ direction. Blue,
red and green balls represent Co, O and Sr atoms respectively. The
top view of the $\mathrm{CoO}$ layer and the $\mathrm{CoO}_{2}$
layer in \textit{b} and \textit{c} are shown in (b) and (c) respectively.
Black arrows in (a) shows the G-AFM spin-ordering. 
In (b) the connected region composed of light green spheres represents the area where a ball with a radius of a $\mathrm{Li^{+}}$ ion can be placed,
so that the corresponding yellow arrow indicates the vacancy channel along the $\left[\bar{1}10\right]$
direction in the $\mathrm{CoO}$ layer.}
\label{fig:str-sco} 
\end{figure}

\textit{Structure} -- The atomic structure of $\SCO$
is shown in Fig.\ref{fig:str-sco}. 
It has orthorhombic symmetry with space group $Pma2$ (28). 
Compared to a $\sqrt{2}\times\sqrt{2}\times4$
perovskite supercell of $\mathrm{SrCoO}_{3}$, the unit cell of $\SCO$ containing eight Sr atoms, eight Co atoms and twenty O atoms
loses four O atoms with two $\mathrm{CoO}$ layers (\textit{b} in Fig.
\ref{fig:str-sco}(a)) appearing. $\mathrm{CoO}$ layers shown in
Fig.~\ref{fig:str-sco}(b) have more spacing than $\mathrm{CoO}_{2}$
layers Fig.~\ref{fig:str-sco}(c). 
Once placing a ball with a radius of $0.76$~\AA, the ionic radius of $\mathrm{Li^{+}}$, into the bulk, 
we then obtained a connected region composed of light green spheres, which is shown in Fig.~\ref{fig:str-sco}(b).
Periodic oxygen-vacancy channel
structures are formed along the $\left[\bar{1}10\right]$ direction
so that they can provide two well-ordered diffusion channels per unit cell marked
in Fig. \ref{fig:str-sco}(b). 
The calculated lattice constant for the orthorhombic
unit cell with eight cobalt atoms, namely $\left(\SCO\right)_{4}$
is $a=5.57$~\AA, $b=5.46$~\AA,  and $c=16.00$~\AA.

\begin{figure}
\includegraphics[width=1\columnwidth]{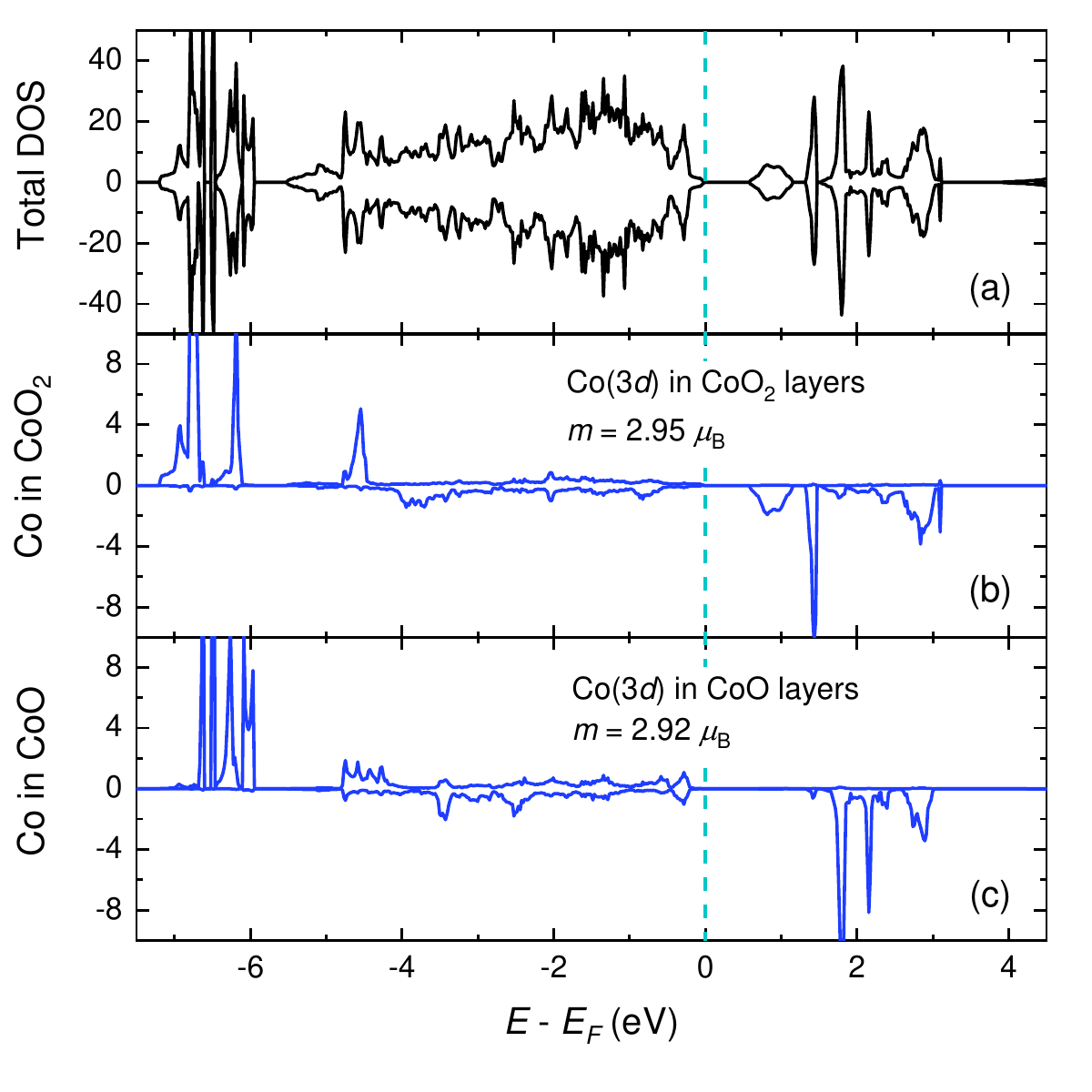}

\caption{In the unit cell of $\left(\SCO\right)_{4}$
under G-AF ordering, (a) the total density-of-state (DOS) and the
projected DOS (PDOS) of Co($3d$) in (b) $\mathrm{CoO}_{2}$ layers
and in (c) $\mathrm{CoO}$ layers. Positive and negative values represent
the spin-majority (spin-up) and spin-minority (spin-down) channels
respectively. The Fermi level is set to zero. The local spin magnetic
moment on Co is also listed.}

\label{fig:dos-sco} 
\end{figure}

\textit{Electronic and magnetic properties} -- The total energy results
show that $\SCO$ has the ground state with
G-type antiferromagnetic (G-AF) or rocksalt-type antiferromagnetism
(AFM) spin ordering, consistent with previous studies\cite{Munoz_2008}. 
G-AF is 68 meV per Co lower in energy than C-type
AFM (C-AF) or column-type AFM, 101 meV lower than A-type AFM (A-AF)
or layered-type AFM and 142 meV lower than ferromagnetic (FM) ordering.
Therefore, the electronic structure of 
$\SCO$ with
G-AF ordering is investigated. According to the density-of-state (DOS)
result shown in Fig. \ref{fig:dos-sco}, a band gap about 0.6 eV indicates
its insulating properties. 
In $\SCO$, O anions and Sr cations have the stable valence state $-2$ and $+2$ respectively, 
so that Co cations have
valence state $+3$ with six $d$ electrons ($d^{6}$) in principle.
Thus, the high-spin state of $\mathrm{Co^{3+}}$ is $S=2$ which
should have $4.0~\mu_{B}$ local magnetic moment with five spin-majority
and one spin-minority $d$ electrons. According to PDOS result in
Fig. \ref{fig:dos-sco}(b)(c), The featured Co($3d$) in both $\mathrm{CoO}_{2}$
and $\mathrm{CoO}$ layers are all below the Fermi level in
the spin-up channel, indicating Co's are indeed in the high-spin state. 
However, the
local moment obtained by DFT is only $2.95~\mu_{B}$. 
According to
the onsite density matrices and the corresponding occupancy of both two types of Co, six orbitals - five in spin-majority
and one in spin-minority - are fully occupied with occupancy closed
to 1, while other four orbitals have occupancy $0.12\sim0.48$, 
indicating unoccupied $3d$ orbitals but hybridized with oxygen ligands under the ligand field. 
Such high Co-O hybridization leads to $\mathrm{Co^{2+}}\underline{L}$ state where the underline $\underline{L}$ refers to a ligand hole\cite{Korotin_1996,Munoz_2008}.
To this end, $d^{6}$ configuration with $S=2$ high spin state on
all $\mathrm{Co^{3+}}$ is still justified.

\section{Lithium-injected $\mathrm{Li}_{x}\SCO$\label{sec:4}}

\begin{figure}
\includegraphics[width=1\columnwidth]{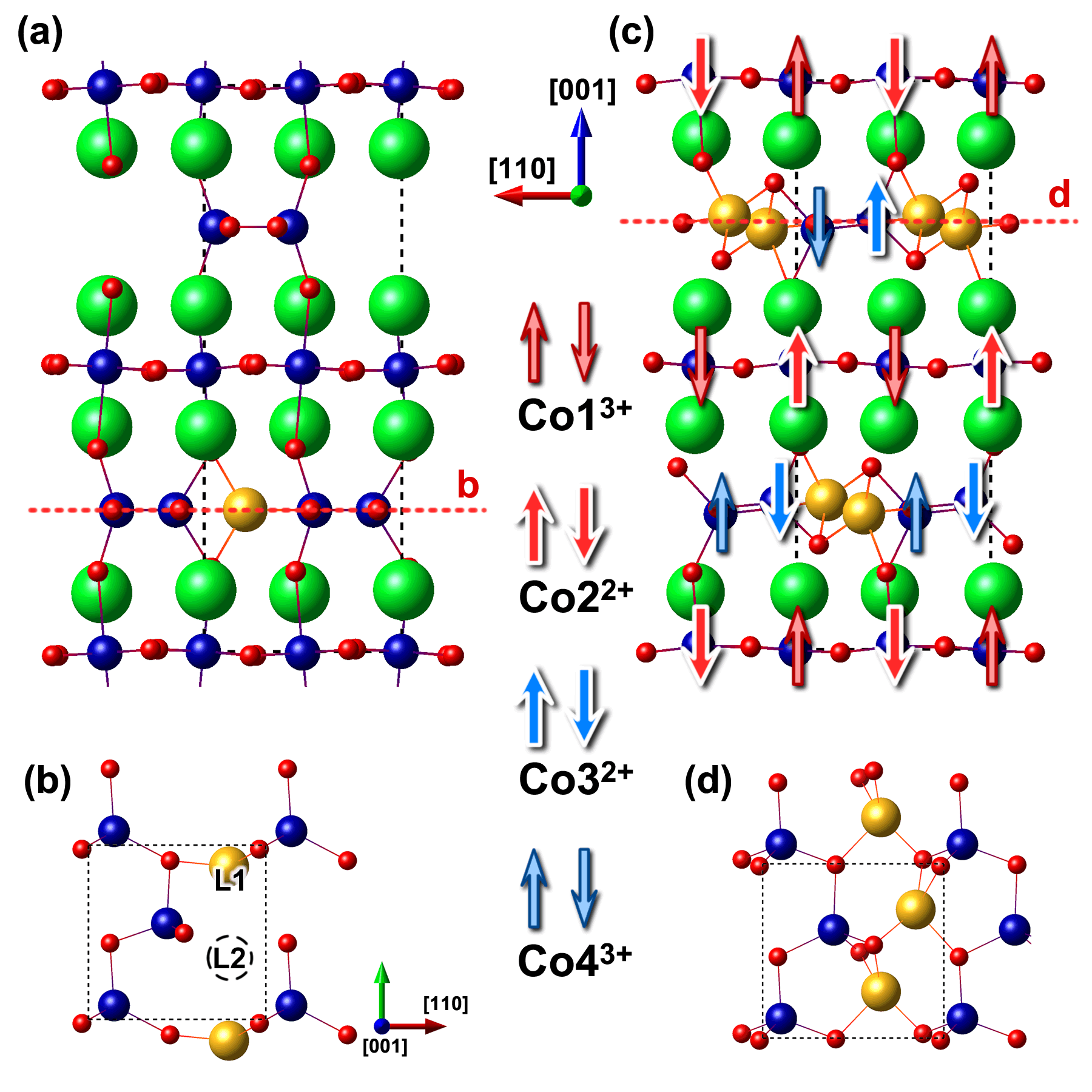}
\caption{Crystal structures of $\mathrm{Li}_{0.25}\SCO$ with
one Li atoms (yellow balls) per unit cell placed in (a) the side view and (b) the top view, and $\mathrm{Li}\SCO$
with four Li atoms per unit cell placed in (c) the side view and (d) the top view.
The dashed lines in side views are the $\mathrm{CoO}$ layers for
top views. 
In (b), two stable sites L1 and L2 are listed.
In (c) the G-AF spin ordering with four non-equivalent cobalt atoms is displayed. 
Co1 and Co4 are in the +3 valence state and Co2 and Co3 are in the +2 valence state.}
\label{fig:str-lsco} 
\end{figure}

\textit{Structures} -- We now investigated the situations where Li atoms are injected into
$\SCO$. 
With one Li placed in the unit cell, the most stable structure of $\mathrm{Li}_{0.25}\SCO$, labeled as L1, 
is shown in Fig.\ref{fig:str-lsco}(a)(b). 
One Li atom is located in the spacing of a $\mathrm{CoO}$ layer, 
at a hollow site in the center of one oxygen-vacancy
channel and have bonds with surrounding three O atoms. 
The Li-O bond-length is $1.86~\AA$, $1.97~\AA$ and $1.99~\AA$ respectively, 
close to the bond-length about $2\AA$~ in lithium oxides\cite{Zintl_1934,Zhuravlev_2010}.
It is distinct from the protonated $\mathrm{H}_{x}\SCO$
where each injected $\mathrm{H}^{+}$ cation is located next to an oxygen
atom with a strong $\mathrm{O-H}$ bond so that $\mathrm{H}^{+}$
cations are not in the same plane with $\mathrm{CoO}$ layers\cite{Lu_2017,Lu_2022}. 
The other meta-stable structure, labeled as L2, where the Lithium ion is located in the middle of two neighboring most stable sites is also identified, shown in Fig.\ref{fig:str-lsco}(b). 
This meta-stable state is about $0.16~\eV$ higher in energy than L1, the most stable one.
Thus, it is obvious that each oxygen-vacancy channel in a $\left(\SCO\right)_{4}$ unit cell can accommodate two Li atoms, 
so that at most four Li atoms can be placed to form $\mathrm{Li}\SCO$, 
shown in Fig.\ref{fig:str-lsco}(c)(d). 
With four Li atoms injected in the unit cell, the lattice constant for $\mathrm{Li}\SCO$
are $a=5.72~\AA$, $b=5.50~\AA$ and $c=16.75~\AA$, only about $8.2\%$ expansion
in volume. This volume expansion is the result of the lattice, which has been internally compressed by the-injected lithium ions, releasing stress.

\begin{table}[b]
\label{tab:lattice}
\caption{The calculated lattice constants for $\mathrm{Li}_{x}\SCO$ with the various concentration of Li ions ($x=0.25,0.5,0.75,1.0$). 
For $x=0.5$, there are three non-equivalent structures labeled L11, L12a and L12b respectively (see main text).}
\begin{ruledtabular}
\begin{tabular}{cccc}
 &$a$ (\AA)&$b$ (\AA)& $c$ (\AA)\\
\hline
$x=0.25$ & 5.66 & 5.50 & 16.02\\
$x=0.5$ (L11)& 5.66 & 5.45 & 16.58\\
$x=0.5$ (L12a)& 5.67 & 5.45 & 16.56\\
$x=0.5$ (L12b)& 5.65 & 5.46 & 16.65\\
$x=0.75$ & 5.69 & 5.51 & 16.73\\
$x=1.0$ & 5.72 & 5.50 & 16.75\\
\end{tabular}
\end{ruledtabular}
\end{table}

We have also investigated the lattice expansion of the structure under different fractional concentrations of lithium ions.
With one Li atom in the unit cell, the volume of $\mathrm{Li}_{0.25}\SCO$ is expand about $2.4\%$.
With three Li atoms intercalated, the volume expansion of $\mathrm{Li}_{0.75}\SCO$ is about $7.8\%$.
For the case of inserting two lithium atoms, there are three non-equivalent configurations. 
The first one is where both lithium atoms are within the same oxygen-vacancy channel (labeled as L11). 
The second one involves two lithium atoms located in different oxygen-vacancy channels and arranged in a staggered pattern (labeled as L12a). 
The third configuration finds each lithium atom occupying a different oxygen-vacancy channel, positioned on the same side (labeled as L12b). 
The volume expansions are $5.0\%$ for L11, $5.2\%$ for L12a, $5.3\%$ for L12b respectively.
The detailed lattice parameters for each structure configurations are listed in Table \ref{tab:lattice}. 
As expected, the volume expansion is essentially linearly related to the concentrations of lithium ions.

\textit{Stability} -- The chemical stability can be confirmed by the formation
energy of $\mathrm{Li}_{x}\SCO$
from bulk $\SCO$ and Lithium metal,
which is defined by 
\begin{equation}
E_{f}=E_{\mathrm{Li}_{n}(\SCO)_{4}}-n E_{\mathrm{Li}}-E_{(\SCO)_{4}}
\end{equation}
where $n=4x$ is the number of Li placed in the $\left(\SCO\right)_{4}$ unit cell and the
total energy for face-centered cubic Li lattice is used for $E_{\mathrm{Li}}$.
The formation energy results for $n=1,3,4$ are $-2.01$, $-5.60$
and $-7.98~\eV$ per unit cell respectively.
Those for all three non-equivalent $n=2$ situations are $-2.93$, $-3.20$ and $-3.74~\eV$ respectively.
They are all negative,
indicating the system reduce energy when $\mathrm{Li}_{x}\SCO$
formed with the injected Li atoms. 
The injected Li atoms can be stabilized in
the oxygen-vacancy channels in all the cases of various
concentration rates. 
Since the formation energy monotonically decreases with the increasing concentration of Li ions, 
one can deduce that once Li ions start to penetrate into $\SCO$, 
the ultimate quasi-static configuration should be full occupancy, 
which means the situation where four lithium ions per unit cell are present.

\begin{figure}
\includegraphics[width=1\columnwidth]{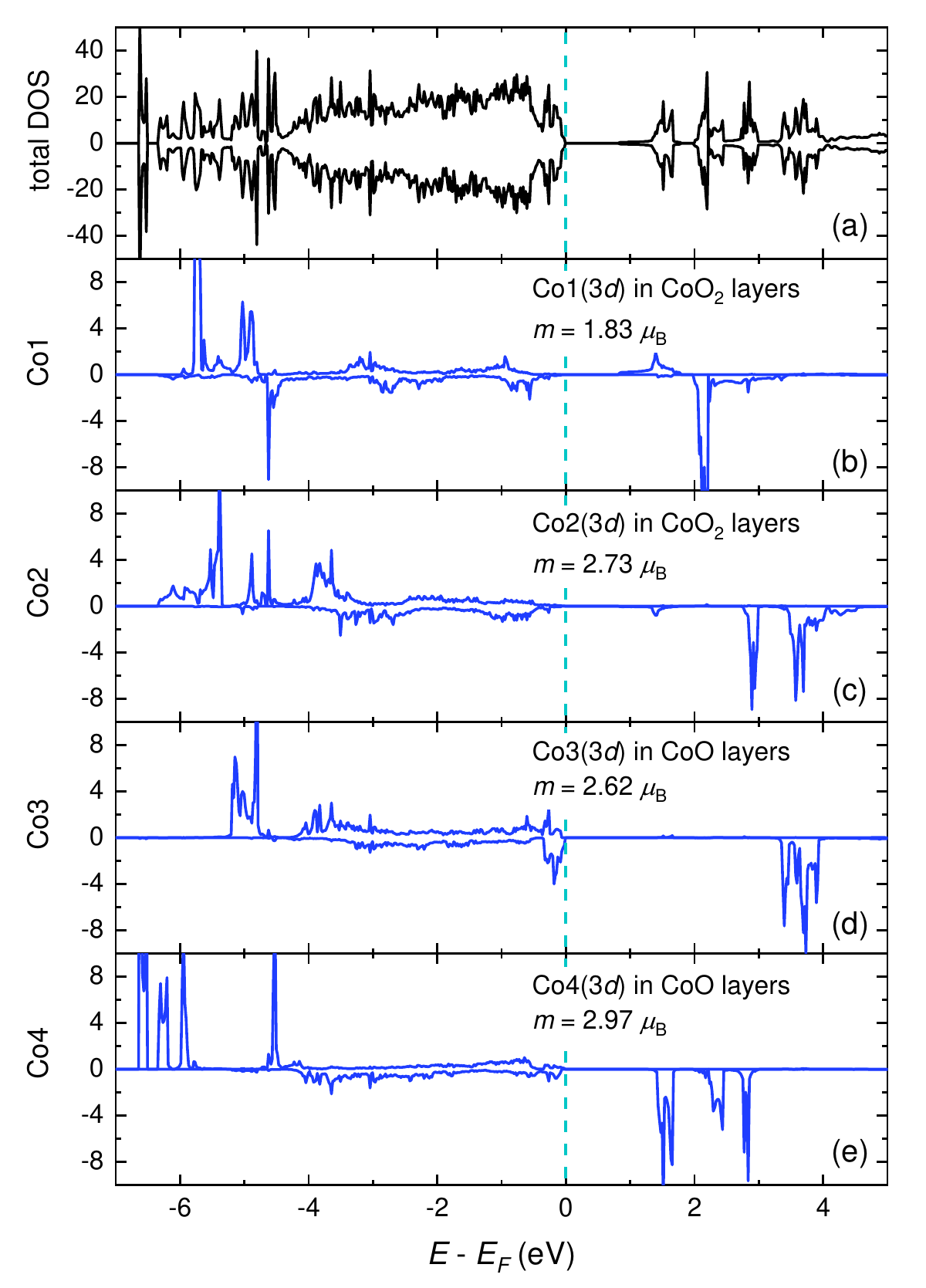}
\caption{In the unit cell of $\left(\mathrm{Li}\SCO\right)_{4}$ in
G-AF ordering, (a) the total DOS and the PDOS of $3d$ orbitals of
Co1, Co2, Co3 and Co4 which is the four Co atoms with positive local
spin magnetic moments in the unit cell. }
\label{fig:dos-l4sco} 
\end{figure}

\textit{Electronic and magnetic properties} -- To investigate how lithium's injection affect the electronic structure and magnetic ordering, we therefore focus on $\mathrm{Li}\SCO$ with all four hollow sites in the unit cell occupied by Li atoms to maximum the effect.
Since each $\mathrm{Li}^{+}$
cation has valence state $+1$, the average valence state of Co is
$+2.5$. In that case, half of Co are $+3$ with $d^{6}$ and half
are $+2$ with $d^{7}$. 
The total energy results still show that the G-AF spin ordering with zero net magnetization
still has the lowest total one. 
G-AF is $11\meV$ per Co lower in energy than C-AF, $12\meV$ lower than A-AF 
and $18\meV$ lower than the FM ordering.
Therefore, in G-AF spin ordering, four Co ions in the unit
cell has spin-up local spin magnetic moment and other four are spin-down. 
The corresponding DOS and PDOS on four
spin-up Co ions labeled Co1, Co2, Co3 and Co4 are shown in Fig.\ref{fig:dos-l4sco}.
The total DOS gives a band gap about 0.8 eV indicating the insulating
properties. The local spin magnetic moments, Co1 and Co2 in $\mathrm{CoO}_{2}$
layers are $1.83~\mu_{B}$ and $2.73~\mu_{B}$ respectively, and those
on Co3 and Co4 in $\mathrm{CoO}$ layers are $2.62~\mu_{B}$ and $2.97~\mu_{B}$
respectively. 
According to PDOS results of Co, Co1(3$d$) in $\mathrm{CoO}_{2}$
layers (Fig. \ref{fig:dos-l4sco}(b)) has unoccupied states above
the Fermi level in the spin-up channel, indicating that Co1 is not
in the high spin state. On the other hand, a sharp peak appears around
$-4.5\mathrm{eV}$ below the Fermi level in the spin-down channel,
indicating that Co1 has more occupied state in the spin-down channel
than $\mathrm{Co^{3+}}$ in $\SCO$. Co2(3$d$)
in $\mathrm{CoO}_{2}$ layers (Fig.\ref{fig:dos-l4sco}(c)) and Co3(3$d$)
in $\mathrm{CoO}$ layers (Fig.\ref{fig:dos-l4sco}(d)) has fully
occupied state in the spin-up channel while in the spin-down channel,
the unoccupied state is located around $3.0\sim4.0\mathrm{eV}$ above
the Fermi level, $1.0~\eV$ higher than the unoccupied state
of $\mathrm{Co^{3+}}$ in $\SCO$. It indicates
that $d$ electrons on Co2 and Co3 in $\mathrm{Li}\mathrm{Sr}_{2}\mathrm{Co}_{2}\mathrm{O}_{5}$
meet stronger on-site Coulomb interaction with more $d$ electrons
than on $\mathrm{Co^{3+}}$ in $\SCO$. PDOS
of Co4 in $\mathrm{CoO}$ layers (Fig.\ref{fig:dos-l4sco}(e)) are
similar to that of $\mathrm{Co}$ in $\SCO$.

We also examined the onsite density matrices and the corresponding occupancies
of all the four types of Co ions. 
For Co1 in $\mathrm{CoO}_{2}$ layers,
four orbitals in spin-majority and two in spin-minority are occupied
with occupancies close to 1, and others have occupancies $0.12\sim0.48$,
indicating unoccupied $3d$ orbitals which only have hybridization bonding-states with the surrounding oxygen ligands.
Therefore, Co1 has +3 valence state and $d^{6}$ configuration with $S=1$
intermediate spin state. For Co2 in $\mathrm{CoO}_{2}$ layers and
Co3 in $\mathrm{CoO}$ layers, five orbitals in spin-majority and
two in spin-minority are occupied with occupancies close to 1, 
so that they are both in +2 valence state and
$d^{7}$ configuration with $S=3/2$ high spin state. 
For Co4 in $\mathrm{CoO}$
layers, five orbitals in spin-majority and one in spin-minority are
occupied with occupancies close to 1,
so that Co4 are in +3 valence state and $d^{6}$ configuration
with $S=2$ high spin state. 
Such metal segregation is caused by the multivalent cobalt ions, 
and happens in other lithium-rich materials\cite{Lin_2019,Boev_2024}
The spin texture of G-AF ordering is therefore displayed in Fig.\ref{fig:str-lsco}(c). 
The insulating properties and the G-AF spin ordering ground state are also confirmed in the electronic structures of $\mathrm{Li}_{x}\SCO$ with $x=0.25, 0.5$ and $0.75$. 
Due to the multivalent cobalt cations,
$\mathrm{Li}^{+}$-injection $\mathrm{Li}_{x}\SCO$
does not affect the insulating properties and the antiferromagnetic
ordering.

\section{Diffusion properties of Li in $\mathrm{Li}_{x}\SCO$\label{sec:5}}

\begin{figure}
\includegraphics[width=1\columnwidth]{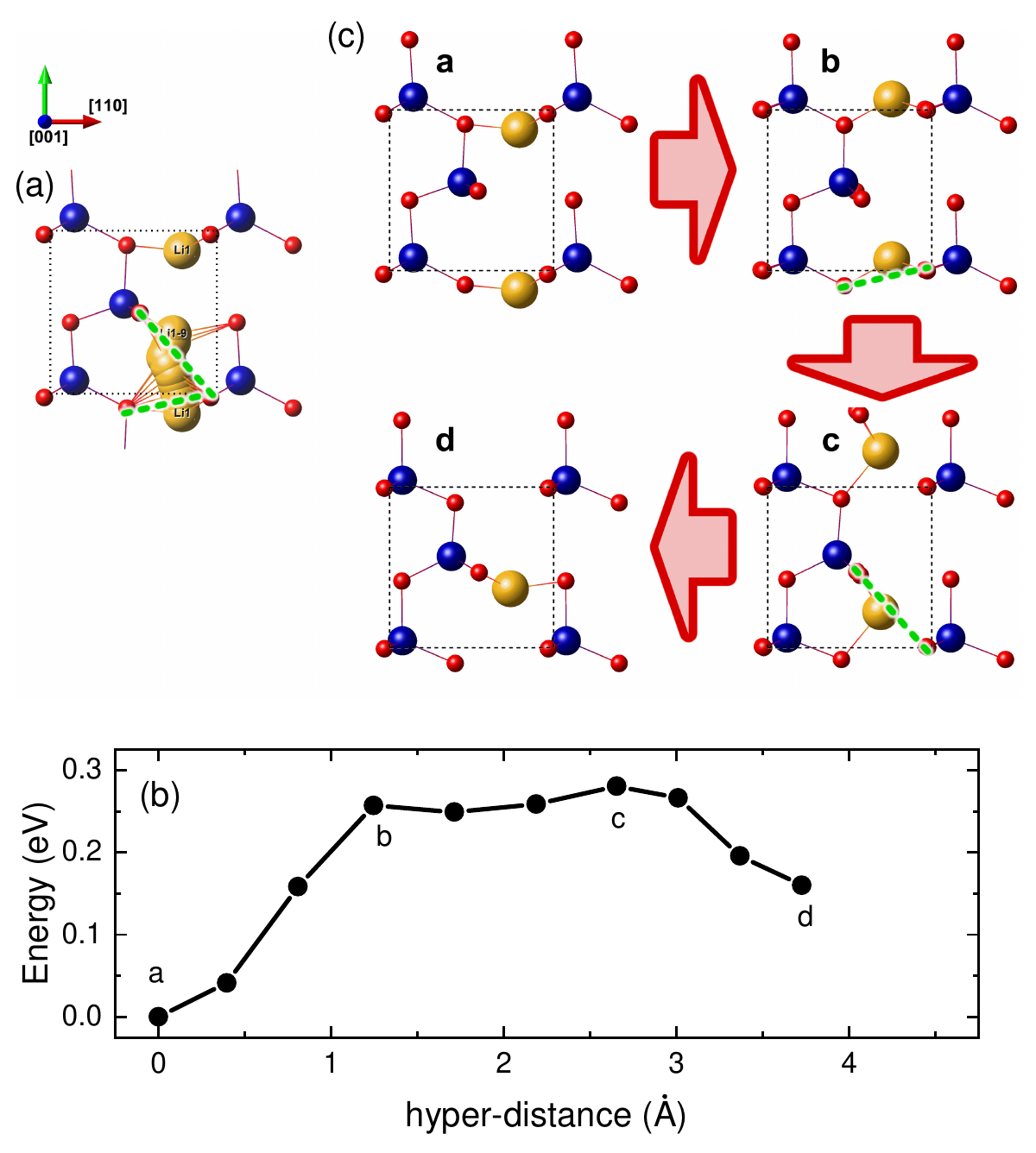}

\caption{cl-NEB results of one Li diffusion in the unit cell of $\mathrm{Li}_{0.25}\SCO$. 
(a) the top view of the migration path of the Li ion diffusing from site L1 to L2 (structure configuration a and d in (c) respectively). 
(b) The total energies of each state relative to the initial state as
a function of hyper-distances between each intermediate image and
the initial state.
(c) the top view of the intermediate images
labeled in (b).
The dashed green lines in (a) and (c) show the two gates composed of two nearby oxygen atoms. 
}

\label{fig:neb-l1} 
\end{figure}

\textit{Migration barriers} -- To investigate the diffusion of $\mathrm{Li^{+}}$ cations in $\mathrm{Li}_{x}\SCO$,
we first calculated the migration barriers of $\mathrm{Li^{+}}$ cations
with various $\mathrm{Li}$ concentration $x$ in $\mathrm{Li}_{x}\SCO$.
We considered two extreme cases.
The first is $\mathrm{Li}_{0.25}\SCO$, which mimics the situation when $\mathrm{Li}^{+}$ cations just begin to insert into $\SCO$ so that very few $\mathrm{Li}^{+}$ cations are injected in the $\SCO$.
To this end, we employed the lattice constant of $\SCO$ for this unit cell.
Therefore, we set the two stable structural configurations L1 and L2
as the initial and ending states.
The other scenario is the opposite, considering the condition of almost full occupancy of Li ions. 
According to the analysis in Section~\ref{sec:4}, 
this situation corresponds to the state with the lowest formation energy, 
which is the quasi-static one after Li ions insertion.
To describe the ionic carriers, 
analogous to the "holes" in p-type electronic semiconductors, 
we employed $\mathrm{Li}_{0.75}\SCO$ with three Li atoms in the unit cell. 
For the lattice constant, 
we used the value of the full occupancy, that is, the $\mathrm{Li}\SCO$ configuration, 
to calculate the migration barrier of the Li ion in the oxygen-vacancy channel which is not fully occupied.
Two different hole positions provide the initial and ending states.
Then, other eight intermediate images on the diffusion pathway are
used to find the lowest migration barriers using
cl-NEB calculations. 

The migration results for $\mathrm{Li}_{0.25}\SCO$ are shown in Fig.\ref{fig:neb-l1}. 
The migration path is straightforward; 
the diffused Li ion starts from the initial state and 
follows an almost straight path along the oxygen-vacancy channel to reach the ending state. 
From Fig.\ref{fig:neb-l1}(a) and (c), it can be observed that the migration path passes through two gates composed of two oxygen atoms each, 
which are the main obstacles during ion migration. 
Consequently, in Fig.\ref{fig:neb-l1}(b), we see that the intermediate states b and c correspond to the states near the two gates, with their potential energy at local maximum. 
Among them, state c has the highest potential energy, 
which is where the migration barrier of the entire migration path is located,
with a magnitude of $0.28~\eV$.

\begin{figure}
\includegraphics[width=1\columnwidth]{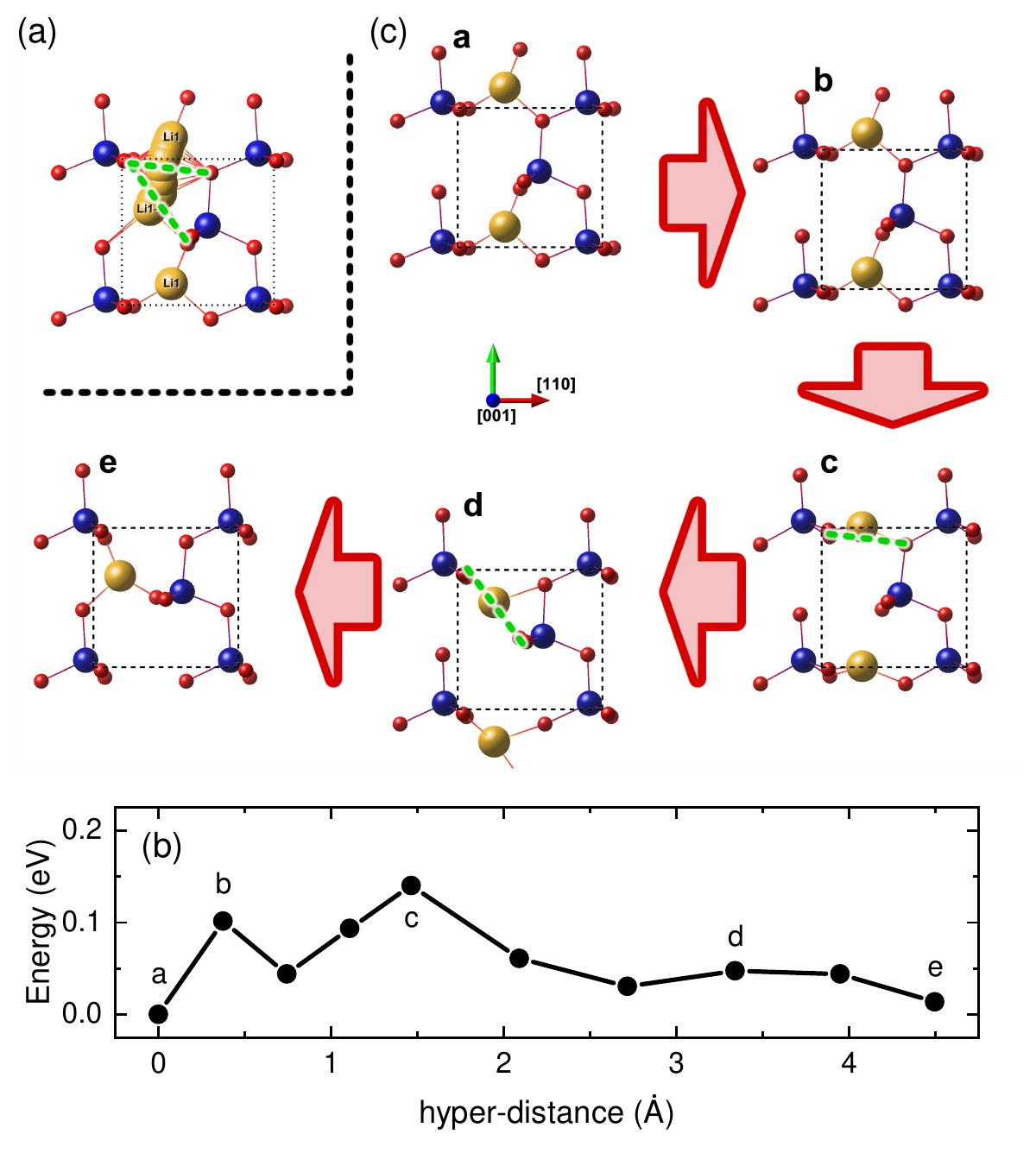}

\caption{cl-NEB results of one Li diffusion in the unit cell of $\mathrm{Li}_{0.75}\SCO$.
(a) the top view of the migration path. 
(b) The total energies of each state relative to the initial state as
a function of hyper-distances between each intermediate image and
the initial state.
(c) the top view of the intermediate images
labeled in (b).
The dashed green lines in (a) and (c) show the two gates composed of two nearby oxygen atoms. 
}

\label{fig:neb-l3} 
\end{figure}

The results for $\mathrm{Li}_{0.75}\SCO$ are shown in Fig.\ref{fig:neb-l3}.
Similar to the case of $\mathrm{Li}_{0.25}\SCO$, the migration path here also essentially takes a straight-line form from the initial state to the ending state. 
Due to the presence of the two gates composed of oxygen atoms, 
intermediate states c and d in Fig\ref{fig:neb-l3}(b) are the two local maxima on the migration path. 
Interestingly, intermediate state b is also a local maximum, which will be analyzed later. 
Among them, the potential energy at the intermediate state c is the highest and is considered to be the migration barrier of $0.17~\eV$ for the entire migration path.
This magnitude of migration
barrier is similar with or even lower than that of $\mathrm{H}^{+}$ diffusion in the
protonated $\mathrm{H}_{x}\SCO$
($\sim0.3~\eV$)\cite{Lu_2022}. 
It is also significantly lower than most of the migration
barriers of the perovskite-type Li-ion solid electrolytes
($0.6\sim0.2~\eV$)\cite{Lu_2021a}.

\begin{figure}
\includegraphics[width=1\columnwidth]{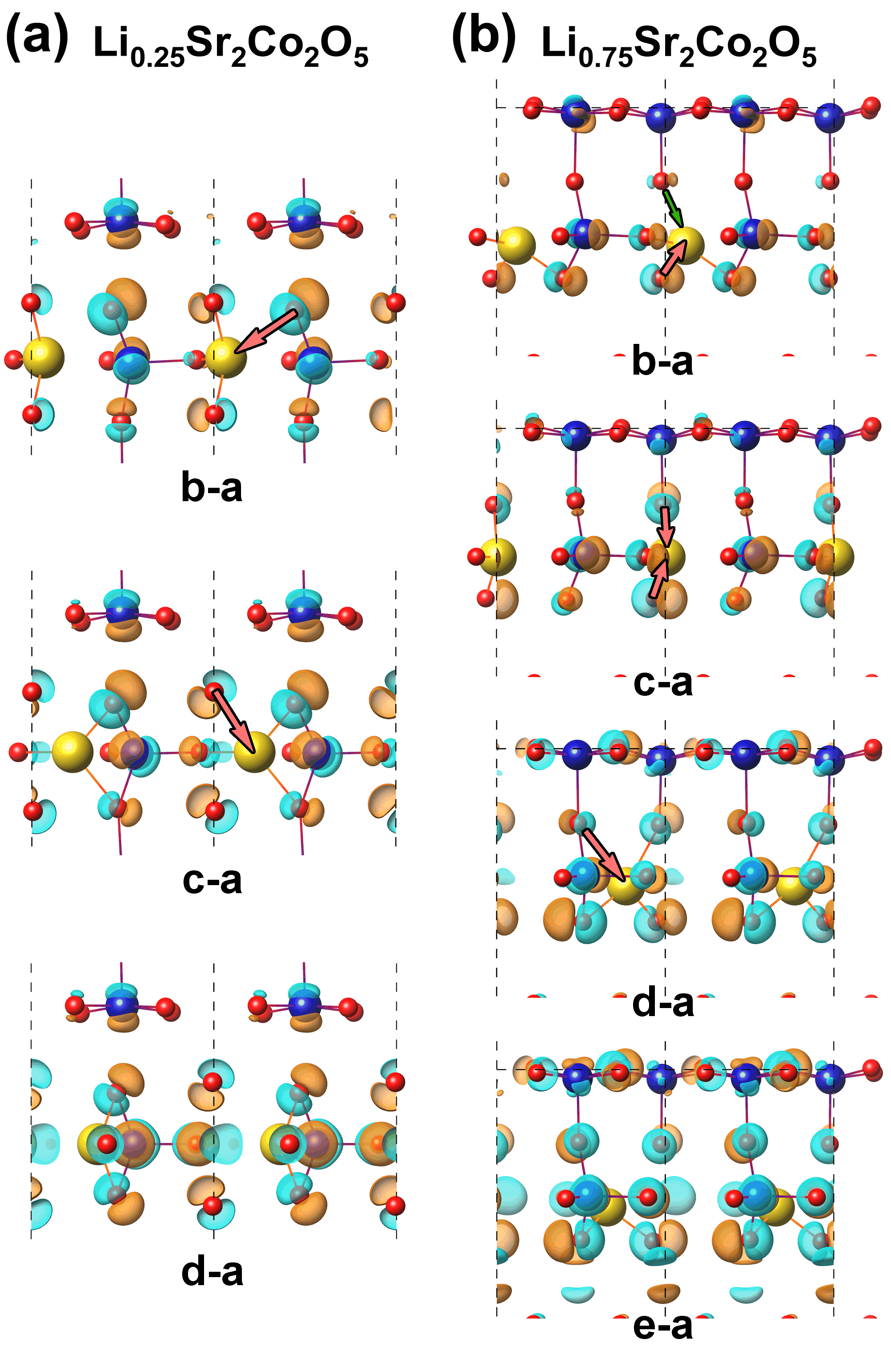}

\caption{In the side views, the charge differences between the initial state (labeled a) and intermediate/ending states (labeled b-d for $\mathrm{Li}_{0.25}\SCO$ and b-e for $\mathrm{Li}_{0.75}\SCO$) for (a) $\mathrm{Li}_{0.25}\SCO$ and (b) $\mathrm{Li}_{0.75}\SCO$. 
The labels corresponds to the label in Fig.\ref{fig:neb-l1} and Fig.\ref{fig:neb-l3}. Orange and cyan isosurfaces correspond to positive and negative values respectively.
The red and green arrows respectively represent the strong interactions that can lower the potential energies during the migration process and the weak interactions that cannot reduce the potential energies.}

\label{fig:neb-chg} 
\end{figure}

To obtain the origin of the low migration barriers of $\mathrm{Li}_{x}\SCO$, we investigated the bonding properties between Li and O during the diffusion.
Fig.\ref{fig:neb-chg} shows the charge difference between the initial state and intermediate/ending states, 
The change of the spatial charge distribution illustrates the change of the chemical bonds during the diffusion,
so that the directions and sizes of isosurfaces can be used to estimate 
the directions and the strength of the interaction between the diffused ion and the neighboring atoms.
In the case of $\mathrm{Li}_{0.25}\SCO$,
the diffused Li ion in the intermediate state b has a strong bond interaction, shown as a red arrow in Fig.\ref{fig:neb-chg}(a), 
though the Li-O distance is about $2.70\AA$ which is much larger than the normal Li-O bonds.
The situation for the intermediate state c is similar to that of intermediate state b, 
where the Li ion has already formed chemical bonds with the subsequent oxygen atoms upon diffusing to that position, 
yet the interaction with the preceding oxygen atom has not been severed, 
as indicated by the red arrow in Fig.\ref{fig:neb-chg}(a).
The newly formed bonds compensate the energy cost of broken bonds, eventually reducing the migration barriers.
In the case of $\mathrm{Li}_{0.75}\SCO$, on the other hand,
the intermediate state b exhibits the opposite effect.
Due to the significant offset of the diffusing Li ion from the center of the oxygen-vacancy channel, 
the distance between this lithium ion and the oxygen atom on one side is relatively large, approximately $2.83\AA$, 
resulting in a noticeably weakened interaction, 
which can be considered as the breaking of the Li-O bond, 
as indicated by the green arrow in Fig.\ref{fig:neb-chg}(b). 
It can be inferred that the energy required for this broken Li-O bond is the main reason for 
the significant increase in the total energy of the intermediate state b. 
In all other respects, 
the situation is similar to $\mathrm{Li}_{0.25}\SCO$, 
meaning that the additional Li-O bond can lower the energy of the intermediate state during the migration.

\textit{Hopping attempt frequency, ionic diffusivity and ion conductivity} -- The ionic diffusivity/conductivity in a solid depends not only on
the migration barriers but also on the pre-exponential factor $a^{2}\nu^{*}$ in Eq.\ref{eq:d-model}. 
According to Vineyard formula\cite{Vineyard_1957, Koettgen_2017},
the hopping attempt frequency can be given by 
\begin{equation}
\nu^{*}=\frac{\prod_{i=1}^{N}\nu_{i}^{0}}{\prod_{j=1}^{N-1}\nu_{j}^{s}}
\end{equation}
where $\nu_{i}^{0}$ and $\nu_{j}^{s}$ are the normal frequencies at the $\Gamma$
point of vibration modes of the initial and saddle point state respectively, 
and $N$ is the total number of phonon modes for the initial state. 
Notice that the number of normal modes in the product of saddle point is one less than that in initial state. 
That is because of the presence of an imaginary frequencies mode at the saddle point state caused by the instability along the diffusion direction. 
This mode needs to be removed from the product. 
The phonon modes at the $\Gamma$ point for
the initial and the saddle point state of both $\mathrm{Li}_{0.25}\SCO$
and $\mathrm{Li}_{0.75}\SCO$
were therefore calculated. 
As a result, the hopping attempt frequencies for $\mathrm{Li}_{0.25}\SCO$
and $\mathrm{Li}_{0.75}\SCO$ are 6.87~THz and 0.95~THz, respectively. 
The lower migration barrier usually leads to a smoother potential energy surface for diffusion, causing the lower force constant and the corresponding vibration frequency. So that the hopping attempt frequency for $\mathrm{Li}_{0.75}\SCO$ is lower than that for $\mathrm{Li}_{0.25}\SCO$.
The hopping distance is regarded
as the half of the lattice constant along $\left[\bar{1}10\right]$
direction, which is $2.75\times10^{-8}$cm for both two phases. 
To this end, for $\mathrm{Li}_{0.25}\SCO$ and $\mathrm{Li}_{0.75}\SCO$,
we can estimate the pre-exponential factor of diffusivity $a^{2}\nu^{*}$,
which is $5.20\times10^{-3}~\cms$ and $0.72\times10^{-3}~\cms$ respectively. 
The hopping attempt frequency can be regarded as the frequency that the diffused ions attempt to climb over the barrier,
so that $\nu^{*}\exp\left(-E_{a}\beta\right)$ is nothing but the hopping frequency from one stable site to its neighbor site with a successful hopping.
At room temperature with $T=300~\K$, $\beta=k_{B}T=38.7~\eV^{-1}$,
the ionic diffusivity $D$ based on Vogele-Tammanne-Fulcher model (Eq.\ref{eq:d-model})
is $1.02\times10^{-7}~\cms$ and $1.00\times10^{-6}~\cms$
respectively. 
If the ionic concentration $n$ is regarded as one $\mathrm{Li}^{+}$ cation per unit cell, the corresponding ionic conductivity $\sigma_{i}=q^{2}nD\beta$
based on Eq.\ref{eq:nernst-einstein} are therefore $1.20\times10^{-3}~\Scm$
and $1.17\times10^{-2}~\Scm$ respectively.
The room temperature diffusivity and conductivity are both superior to most of the
perovskite-type Li-ion solid electrolytes 
($10^{-4}\sim10^{-3}~\Scm$ for bulk conductivity)\cite{Lu_2021a}, 
garnet-type solid-state electrolytes ($10^{-4}\sim10^{-3}~\Scm$)\cite{Wang_2020}, and sulfide solid electrolytes ($10^{-7}\sim10^{-3}~\Scm$)\cite{Manthiram_2017,Wu_2023a}.

The above estimates of diffusivity and conductivity are based on an assumption 
that diffusion over each energy barrier is adiabatic, 
and the mechanical energy of the migrating lithium ion 
is conserved from the initial state to the final state. 
However, when there are several local maxima of the potential energy along the diffusion path, 
it is possible that after overcoming each local barrier, 
the lithium ion may reset its kinetic energy through scattering due to thermal fluctuations.
Here, we consider a strong scattering limit where the lithium ion resets its kinetic energy after surmounting each local barrier. 
Thus, the diffusion path is divided into independent segments, with each segment's diffusivity calculated separately. 
For $\mathrm{Li}_{0.25}\SCO$, 
the original diffusion path from point a to d in Fig.\ref{fig:neb-l1}(b)  
is split into two segments for overcoming the local barriers at b and c.
The diffusion distances for these segments are $1.95\AA$ and $0.80\AA$, respectively,
and the corresponding relative barriers are $0.26\eV$ and $0.03\eV$, respectively. 
Using the previously mentioned method, 
the hopping attempt frequencies are 4.60~THz and 5.61~THz respectively. 
Thus, the resulting diffusivity in the room temperature is $0.75\times10^{-7}~\cms$ and $1.12\times10^{-4}~\cms$, respectively.
Considering that the two segments are connected in series, 
with the second segment's diffusivity being much larger than the first, 
the overall conductivity is determined primarily by the first segment. 
Consequently, the resulting conductivity is calculated to be $0.88\times10^{-3}~\Scm$.

In the case of $\mathrm{Li}_{0.75}\SCO$, 
the whole diffusion path can also be divided into three segments, 
corresponding to three local barriers at b, c and d in Fig.\ref{fig:neb-l3}. 
The relative barriers are $0.10\eV$, $0.10\eV$ and $0.02\eV$ respectively. 
The hopping attempt frequencies obtained are 0.39~THz, 0.95~THz and 7.09~THz respectively.
The corresponding diffusivity in the room temperature is $1.65\times10^{-7}~\cms$, $3.51\times10^{-6}~\cms$ and $3.08\times10^{-4}~\cms$ respectively. In the case of series connection, the overall conductivity is determined primarily only bey the first and second segments, which is about $1.85\times10^{-3}~\Scm$

Under the strong scattering limit, the calculated ionic conductivity for both cases has decreased, but it remains within the same order of magnitude. 
The actual situation would be between the cases of the adiabatic approximation and strong scattering.
Considering that the diffusion length in a diffusion period is only $2.75~\AA$, 
the actual conductivity should be closer to the results of the adiabatic approximation.



\section{Conclusions}

In summary, We investigated the structural, electronic and magnetic properties of brownmillerite $\mathrm{Li}_{x}\SCO$ and confirmed the oxygen vacancy channels, the insulating property and the G-AF spin ordering ground state.
At most four $\mathrm{Li}^{+}$ cations can be stabilized in the center of the oxygen vacancy channels of an orthorhombic $\mathrm{\left({Sr}_{2}{Co}_{2}{O}_{5}\right)_{4}}$ unit cell to form $\mathrm{Li}_{x}\SCO\left(x=0.0\sim1.0\right)$. 
The G-AF spin ordering and insulating property is still remained with the lithium's injection.
By employing cl-NEB calculations, we obtained the migration barriers of $\mathrm{Li}_{x}\SCO$ for $x=0.25$ and $0.75$ as $0.28~\eV$ and $0.17~\eV$, respectively. 
After obtaining the hopping attempt frequencies via Vineyard formula, 
we therefore obtained the corresponding diffusivity and conductivity at the room temperature as $1.02\times10^{-7}~\cms$ and $1.20\times10^{-3}~\Scm$ for $x=0.25$ respectively, 
and $1.00\times10^{-6}~\cms$ and $1.17\times10^{-2}~\Scm$ for $x=0.75$ respectively. 
The high ionic diffusivity and ion conductivity indicate brownmillerite $\SCO$ as a promising super-ionic conductor. 
Because the structure of brownmillerite-type is very similar to that of the perovskite-type, 
$\mathrm{Li}_{x}\SCO$ inherits the advantages of the perovskite-type solid electrolyte,
which include high Li-ion conductivity, low electron conductivity, and wide electrochemical window.
In particular, due to the exclusive occupation of the oxygen vacancy channels by Li ions in $\mathrm{Li}_{x}\SCO$, 
it can address the issue of poor stability against lithium metal faced by perovskite-type solid electrolytes\cite{Lu_2021a, Wu_2023}.
We expect that $\mathrm{Li}_{x}\SCO$ can be compatible with the application of perovskite-type solid electrolytes in all-solid-state lithium batteries.

\begin{acknowledgments}
We are grateful for fruitful discussions with Prof. Pu Yu. 
This work was financially supported by the National Natural Science
Foundation of China (12274309). Analysis of magnetic states done at UNH was supported by the U.S. Department of Energy, Director, Office of Science, Office of Basic Energy Sciences, Division
of Materials Sciences and Engineering under Contract No. DE-SC0020221.
The cl-NEB calculations were performed at Shaheen II in King Abdullah University of Science and Technology (KAUST).
\end{acknowledgments}

\bibliographystyle{apsrev4-2}
\bibliography{ref_1}

\end{document}